\documentclass[runningheads]{llncs}
\usepackage[T1]{fontenc}
\usepackage{graphicx}
\usepackage[most]{tcolorbox}
\usepackage{xcolor}
\usepackage{enumitem}
\usepackage{tabularx}
\usepackage{booktabs}
\usepackage{colortbl}
\usepackage{listings}
\usepackage{changepage}
\usepackage{multirow}

\begin{document}
\title{Metadata-driven Table Union Search: Leveraging Semantics for Restricted Access Data Integration}
\titlerunning{Metadata-driven Table Union Search}
%
\author{Margherita Martorana\inst{1}\orcidID{0000-0001-8004-0464} \and
Tobias Kuhn\inst{2,3}\orcidID{0000-0002-1267-0234} \and
Jacco van Ossenbruggen\inst{3}\orcidID{0000-0002-7748-4715}}
\authorrunning{M. Martorana et al.}
%
\institute{Vrije Universiteit Amsterdam, The Netherlands
\email{m.martorana@vu.nl}}

\maketitle              
\begin{abstract}

    

Over the past decade, the Table Union Search (TUS) task has aimed to identify unionable tables within data lakes to improve data integration and discovery. While numerous solutions and approaches have been introduced, they primarily rely on open data, making them not applicable to restricted access data, such as medical records or government statistics, due to privacy concerns.  Restricted data can still be shared through metadata, which ensures confidentiality while supporting data reuse. This paper explores how TUS can be computed on restricted access data using metadata alone. We propose a method that achieves 81\% accuracy in unionability and outperforms existing benchmarks in precision and recall. Our results highlight the potential of metadata-driven approaches for integrating restricted data, facilitating secure data discovery in privacy-sensitive domains. This aligns with the FAIR principles, by ensuring data is Findable, Accessible, Interoperable, and Reusable while preserving confidentiality.


\end{abstract}
\section{Introduction}
Tabular data is one of the most common formats for data storage and sharing, widely used in industries and research domains \cite{nguyen2019mtab}. With the ever-growing volume of available data, several initiatives have been established, such as the Center for Open Science\footnote{\url{https://www.cos.io}} and the Linked Open Data Cloud\footnote{\url{https://www.lod-cloud.net}}, to promote data discovery and reuse through recommendation systems and search engines \cite{foster2017open}. Despite these efforts, searching for tables on the Web and within repositories remains an active and evolving research area \cite{sarma2012finding} \cite{limaye2010annotating} \cite{zhu2016lsh} \cite{miller2018making}.

The Table Union Search (TUS) problem, first introduced by \cite{nargesian2018table} in 2018, focuses on identifying ``the k tables that are most likely to be unionable with a given search table''. Since then, several methods and benchmarks have been proposed to improve table union search \cite{bogatu2020dataset} \cite{fan2022semantics} \cite{khatiwada2023santos} \cite{pal2024alt}. However, these approaches are highly dependent on having access to the underlying data. For restricted access data, such as medical records or official government statistics, this assumption does not apply due to confidentiality and privacy restrictions that prevent open sharing. Restricted access data, while not openly available, remains vital for research, particularly in fields such as healthcare and social sciences. In recent years, there has been growing interest in making such data more Findable, Accessible, Interoperable, and Reusable (FAIR) \cite{wilkinson2016fair}. Previous works have shown that high-quality and semantically rich metadata can facilitate the discovery, reuse, and integration of restricted access data \cite{martorana2022aligning} \cite{martorana2023advancing} \cite{martorana2024column}. In fact, metadata can provide descriptive information about the dataset without exposing sensitive information, ensuring confidentiality while enabling data discovery and integration. 

In this work, we present a method to perform Table Union Search on restricted access data, using metadata only. To illustrate the applicability of our approach, we introduce the following use case.

\begin{tcolorbox}[colback=blue!10, colframe=blue!40, rounded corners, boxrule=0.5pt]
\textbf{Use Case 1.} A researcher is looking for data to answer a complex question, such as ``How does the recovery time from health condition X differ across regions in The Netherlands, and how is it influenced by the level of patient education?''. To answer this, the researcher needs data from healthcare providers, as well as data on education levels, which may be collected by government agencies or educational institutions. The type of data is both complex and confidential, and comes from different providers. Using a metadata portal, the researcher can discover and search for relevant datasets based on metadata alone. However, current systems cannot assess unionability, whether datasets can be meaningfully combined, leaving the researcher unsure if the required datasets are available and useful to answer their research questions.
\end{tcolorbox}

Our method also has broader applications in the domain of open data and web table search systems. Existing table search algorithms, such as Google Dataset Search \cite{brickley2019google}, rely on metadata for retrieval and typically use basic keyword matching to search metadata records. However, metadata is often incomplete, inconsistent, or of low quality \cite{adelfio2013schema} \cite{farid2016clams} \cite{nargesian2019data} \cite{zhang2020finding}. Despite these challenges, certain information, such as column headers, can still be automatically extracted and used to enrich metadata. These enhancements can significantly improve the accuracy of searches for unionable tables. To demonstrate this broader applicability, we introduce Use Case 2.

\begin{tcolorbox}[colback=blue!10, colframe=blue!40, rounded corners, boxrule=0.5pt]
\textbf{Use Case 2.} A researcher is looking for data to answer a question such as ``How does unemployment rate affect the availability of educational programs in different countries?''. Although the data is publicly available, it is hard to find because the datasets might be spread across different sources and data providers. Additionally, the dataset retrieval systems used by the researcher cannot automatically determine whether the required datasets can be meaningfully combined. Although the researcher could manually examine each dataset, this process is time consuming and labor intensive.
\end{tcolorbox}

Although the primary focus of our work is on the restricted access data domain, it can also benefit the open data domain. By addressing these challenges, our method leverages semantic technologies and rich metadata to facilitate the discoverability and usability of open and restricted access data. The main research questions that we aim to answer in this work are the following:
\begin{itemize}[label=\textbullet, left=0pt, itemsep=5pt]
    \item[]\textit{How can we perform Table Union Search on restricted access data?}
    \item[]\textit{How can semantic technologies be leveraged while performing Table Union Search on restricted access data?}
    \item[]\textit{How do methods for Table Union Search on restricted access data compare to existing benchmarks using open data?}
\end{itemize}

The key contribution of this paper is a method for computing table unionability for restricted access data, solely using only metadata information. This approach allows us to assess whether datasets can be meaningfully combined while preserving confidentiality.

\section{Background}
As follows, we firstly present the state-of-the-art methods in table union search, then we discuss the challenges associated with restricted access data, and highlight the recent advances in metadata-driven data analysis and integration. Finally, we define the method addressed in this research: Metadata Union Search. 

\subsection{Table Union Search}
Table union search involves identifying a set of candidate tables \textit{C} within a data lake that can be unioned with a given query table \textit{Q} \cite{nargesian2018table}. The concept was first formalized by \cite{nargesian2018table}, who introduced a labeled benchmark for Table Union Search (TUS) on open data. They considered a candidate table unionable with a query table if a subset of their columns were unionable, determined by a combination of statistical tests: i.e. value overlap, semantic overlap, and word embedding similarity of column values. To construct the benchmark, they horizontally and vertically sliced 32 large seed tables to create smaller unionable tables. However, this approach had limitations: the variety of semantics was constrained by the small number of seed tables, and column headers were not included in the ground truth and benchmark data. 

This initial approach was extended by \cite{bogatu2020dataset} with the D\textsuperscript{3}L method, which incorporated three additional attributes for assessing unionability: column header similarity, numerical value distribution, and regular expression similarity. These enhancements aimed to improve the precision of unionability predictions by leveraging a broader set of features. Subsequent work, STARMIE \cite{fan2022semantics}, introduced a new approach employing a contrastive learning approach to better capture the context of the entire query and candidate tables. This approach aimed to leverage contextual relationships within tables, moving beyond individual column similarities. Another significant advancement came with SANTOS \cite{khatiwada2023santos}, which evaluated both column similarity and binary relationships between columns. By incorporating these relationships, SANTOS improved the understanding of table context and avoided incorrect unions of tables that had similar column values but different semantics. However, the SANTOS benchmark was limited, providing only 80 human-labeled samples in the ground truth.

The most recent development in table union search is the ALT-gen benchmark, introduced by \cite{pal2024alt}. ALT-gen uses generative AI to create benchmark datasets for table union search. This benchmark includes tables generated from 50 topics (e.g., medicine, archaeology, sociology) and provides detailed validation against hallucinations. The benchmark consists of 50 query tables and 1,000 candidate tables (500 unionable and 500 non-unionable). ALT-gen is the most detailed benchmark to date, offering complete access to query and candidate tables, including their topic information, column headers, and a fully labeled ground truth of unionable and non-unionable tables.

Despite these advances, current table union search methods and benchmarks rely heavily on access to the underlying data for assessing unionability. In this work, we leverage the ALT-gen benchmark to explore how can table union search be performed for restricted access data. Specifically, we investigate how metadata information alone can be used to perform the table union search task, addressing challenges in contexts where direct access to the full data is not possible.

\subsection{Data Integration through Metadata-Driven Approaches}
Recent advances in data sharing and reuse have highlighted the critical role of metadata in enabling access to restricted datasets, particularly those containing sensitive information \cite{martorana2022aligning}. Efforts have emerged to address challenges related to privacy concerns, such as individual-level data often found in health records or demographic datasets, while still facilitating scientific exploration and collaboration. Open Government Data (OGD) portals, such as the Central Bureau for Statistics Netherlands (CBS)\footnote{\url{https://www.cbs.nl/}}, the U.S. Government's Open Data\footnote{\url{https://data.gov/}}, and Canada's Open Government Portal\footnote{\url{https://search.open.canada.ca/data/}}, are examples of platforms designed to provide aggregated population data for research and public use. Despite their utility in domains like scientific research and software development \cite{begany2021open}, these portals do not directly address the challenges around individual-level data, due to its confidential nature. For instance, datasets containing sensitive information like medical records or personally identifiable information (PII) remain inaccessible in their raw form.

To address this, solutions such as the Personal Health Train \cite{deist2020distributed} have been proposed, which allows algorithms to be executed directly at the location where the data is stored, ensuring that users can analyze data without accessing it. However, this approach does not resolve the challenge of discovering and understanding restricted access datasets beforehand, and without detailed structural and contextual information, it remains difficult to design meaningful analysis algorithms. Metadata repositories have emerged as a possible solution to this problem. Portals like the Open Data Infrastructure for Social Science and Economic Innovation (ODISSEI)\footnote{\url{https://odissei-data.nl}} focus on storing metadata for restricted access datasets, such as those from CBS. These platforms allow researchers to explore dataset availability and structure before starting lengthy and costly processes to request access. High quality metadata has proven to be the most effective way to make restricted datasets more Findable, Accessible, Interoperable, and Reusable (FAIR) \cite{wilkinson2016fair}, as highlighted in previous studies \cite{martorana2022aligning}. Rich metadata standards, such as the Dataset-Variable Ontology (DSV) \cite{martorana2023advancing}, enable the annotation of restricted datasets even at the column level, capturing semantic and structural information without exposing confidential data.

Recent advancements in metadata-driven approaches have also investigated the process of semantically enriching the metadata of restricted access datasets \cite{martorana2024zero} \cite{martorana2024column}. For instance, linking column headers to controlled vocabularies, such as the CESSDA Topic Vocabulary\footnote{\url{https://vocabularies.cessda.eu/vocabulary/TopicClassification?lang=en}} or the DBpedia\footnote{\url{https://www.dbpedia.org/about/}} properties ontology, allows researchers to incorporate rich semantic context into metadata annotations. These efforts can enhance the FAIRification of restricted datasets, promoting interdisciplinary collaboration and improving dataset discovery across domains \cite{vlachidis2021semantic} \cite{sasse2022semantic}. The annotation of column-level metadata has been shown to facilitate the retrieval and use of sensitive datasets, including medical records and microdata, by preserving privacy while adding valuable context \cite{dugas2016odmedit} \cite{magagna2021adopt} \cite{jonquet2023common} \cite{razick2014egenvar}.

The concept of ``Dataless Tables'' has been recently introduced by \cite{martorana2024column}, highlighting the potential of metadata-driven approaches. Dataless Tables represent restricted access datasets whose underlying data is inaccessible, but still provide detailed metadata, including structural information, summary statistics, and semantic annotations. These tables offer a novel way to enable data discovery, integration, and analysis while adhering to strict privacy requirements. By leveraging such metadata, researchers can engage with restricted datasets in a meaningful way without compromising confidentiality.

Metadata-driven analysis is not only crucial for restricted access datasets but it can also play a significant role in dataset retrieval systems, which often rely on metadata and keyword search for indexing and retrieval. The integration of semantic metadata enrichment and controlled vocabularies has the potential to benefit both restricted and openly accessible datasets, facilitating interoperability and reuse across diverse applications.

\subsection{Metadata Union Search}
We define the concept of \textbf{Metadata Union Search} as the process to compute unionability for restricted access datasets (i.e. \textit{dataless tables}). Traditional benchmarks have assessed unionability by analyzing the semantic and value overlap between the underlying data of a query dataset and candidate datasets within a data lake. However, in scenarios where data access is restricted, this approaches are not feasible. Instead, our methodology focuses exclusively on metadata information to evaluate unionability.

We define unionability as the semantic similarity between the metadata of the query dataset and the candidate datasets, evaluated specifically at the column level. Prior research has demonstrated that column-level metadata can provide sufficient detail to describe the semantic of a dataset \cite{martorana2023advancing} \cite{martorana2024zero} \cite{martorana2024column}, such as by capturing structural, contextual and semantic elements without exposing sensitive information. In the sections that follow, we detail our approach to enriching metadata with semantic technologies, and evaluate our MUS method against other available approaches.

\section{Experimental Design}
In this section, we describe the experimental design of our study, which focuses on computing table unionability using solely metadata information. Our approach is tested in multiple settings that involve different types of metadata enrichment, which will be described in more detail below. We define two distinct scenarios for evaluating unionability:

\begin{enumerate}[left=0pt, itemsep=5pt]
    \item \textbf{Topic Dependent (TD)}: In this scenario, a researcher is searching for unionable tables within the same domain (or topic) as the query table, stored in a data lake that provides filtering capabilities based on domain. The selection of candidate tables for unionability is explicitly constrained by the topic of the query table, ensuring domain-specific results.
    \item \textbf{Topic Guided (TG)}: In this scenario, no domain-specific filtering is available. Instead, the researcher aims to explore unionability across tables from various domains and topics. Here, the topic of the query table serves as a guiding factor, influencing the evaluation of unionability while allowing candidate tables from different domains to be considered.
\end{enumerate}

These two scenarios provide the basis for our approach and are used to test how metadata-based unionability may be applied to both domain-specific and general searches.

\subsection{Data Preparation}
The data used for this experiment was sourced from the ALT-gen benchmark \cite{pal2024alt}, which consists of 1050 tables in total. This includes 50 query tables and 1000 candidate tables across 50 different topics, which each topic containing one query table and 20 candidate tables. The data is provided in CSV format. 

To process the data, we used a Python script to extract the metadata from the CSV files and convert it into RDF format. This conversion follows the DataSet-Variable Ontology (DSV) proposed in \cite{martorana2023advancing}. Since our research focuses on restricted access data, we created metadata that captures column-level information, such as column headers, but excluded any underlying data values. 

Once the metadata was extracted and converted into RDF format, it was enriched using the Python RDFLib library\footnote{\url{https://rdflib.readthedocs.io/en/stable/}}. This allowed us to further process and enhance the metadata with semantic enrichments. 

After the metadata was prepared, we divided it into testing and evaluation sets. We randomly selected 40\% of the topics defined by ALT-gen to be included in the test set, along with their corresponding query and candidate tables. The remaining 60\% were used for evaluation. Unlike the original ALT-gen benchmark, which assigns binary values (0 or 1) to indicate whether the datasets are unionable or not, we calculate unionability based on semantic similarity. The test set was used to determine a threshold for cosine similarity, where pairs of query and candidate tables with cosine similarity above the threshold were considered unionable. 

All codes and data used in this work is available at \footnote{\url{https://github.com/ritamargherita/MetaUnionSearch}}.

\subsection{Metadata Enrichment}
The metadata enrichment process was carried out in two different ways, each with the aim of enhancing the semantic representation of the datasets. 
\textbf{Firstly}, we enriched the metadata by incorporating \textit{semantic data types}. This was achieved using the ``Sherlock model'' from \cite{hulsebos2019sherlock}, which is a deep learning model designed to detect semantic data types from tabular data. Each CSV file from the ALT-gen benchmark was processed with the Sherlock model, which identified a semantic data type for each column. These semantic data types were then added to the column-level metadata using the \verb|dcterms:type| property.
\textbf{Secondly}, we annotated each column-level metadata with DBpedia properties. This annotation leveraged a Large Language Model (LLM) approach, following the methodology of \cite{martorana2024column} \cite{martorana2024zero}. In this process, we began by computing the embeddings for both the column headers and the DBpedia properties, using the ``all-MiniLM-L6-v2'' model and the SentenceTransformer library in Python. Cosine similarity score was calculated between the embeddings, and for each column header the ten most similar DBpedia properties were identified. Then, we used the OpenAI API and the GPT-4o-mini model to select for each column the most suitable DBpedia property from this subset. The selected property was added to the column-level metadata using the \verb|dsv:columnProperty| property. 

\subsection{Computing Unionability on Metadata}
To compute table unionability for restricted access data, we measure the semantic similarities between the metadata of the query and the candidate tables. This is achieved by calculating the cosine similarity between the column-level metadata of the respective tables. 

Similarly to finding the DBpedia properties per column header, explained above, the process of computing the unionability from solely metadata information starts with creating embeddings. The embeddings, again, were generated using the ``all-MiniLM-L6-v2'' model and the SentenceTransformer library in Python. The embeddings were created for each term within the column-level metadata For example, in the case of enrichment with DBpedia properties, separate embeddings were created for the column header label and the DBpedia property label. These embeddings were then combined by summing and averaging the vectors. 

The approach is slightly different based on the two experimental scenarios. In the TD scenario, the dataset topic was not embedded, and only column-level metadata and its enrichments were used to compute similarity. In the TG scenario, the dataset topic was included in the embeddings alongside the column-level metadata and other enrichments. To account for its significance, the topic embedding was assigned a weight of 25\%, while the remaining 75\% of the embedding consisted of the other column-level metadata (i.e. column header label, DBpedia properties and semantic data types). 

After computing the embeddings, cosine similarity was used to identify the candidate datasets that were most semantically similar to the query dataset. With this approach, we aim to provide a scalable and effective way to compute unionability for restricted access data, based solely on metadata. 

\subsection{Evaluation}
The groundtruth for unionability was also sourced from the ALT-gen benchmark. The groundtruth provides annotation for each query dataset and its candidate datasets, indicating whether the pair is unionable (1) or not (0). 

In our approach, unionability is computed based on embeddings and semantic similarities between metadata, which required defining a threshold to determine when a dataset pair is considered unionable. To establish this threshold, we calculated embeddings and cosine similarity scores for the test set under each scenarios. These scores were then compared to the groundtruth annotation from ALT-gen. To derive the optimal threshold, we used the Youden's statistic, which balances sensitivity and specificity to maximise the overall accuracy of the classification. This threshold was subsequently applied to the evaluation set: query and candidate pairs with cosine similarity scores above the threshold were classified as unionable. 

The final unionability results for the evaluation set, calculated with the threshold, were compared against the groundtruth to asses the accuracy of our approach. Additionally, we evaluated our methodology against alternative unionability methods previously tested with the ALT-gen benchmark, namely D\textsuperscript{3}L\cite{bogatu2020dataset}, SANTOS\cite{khatiwada2023santos} and Starmie\cite{fan2022semantics}. The comparison was carried out using the mean average precision at k (MAP @ k), recall at k (R @ k), and precision at k (P @ k), where \verb|k = 10|. These metrics allowed us to compare and evaluate the effectiveness of our approach in computing table unionability against other available methodologies.

\section{Results}
In the following section, we present the results of our experiments, starting with a summary statistic from the creation and enrichment of the metadata. We then provide the results of the unionability computations and a comparison with other available methods.

\subsection{Metadata Creation and Enrichment}
The first step in our experimental design was to create metadata from the CSV files of the ALT-gen benchmark. The dataset consisted of 1,000 candidate tables and 50 query tables, each corresponding to one of the 50 topics. For our experiments, we divide the data into a test set (40\%) and an evaluation set (60\%). This means that the test set contains 20 query datasets and 400 candidates, while the evaluation set included 30 queries and 600 candidates. To process and enrich metadata, we utilized a Python script and the RDF-lib library, following the DSV ontology to describe the metadata of restricted access datasets. 

\begin{figure}[htbp]
    \centering
    \includegraphics[width=0.8\textwidth]{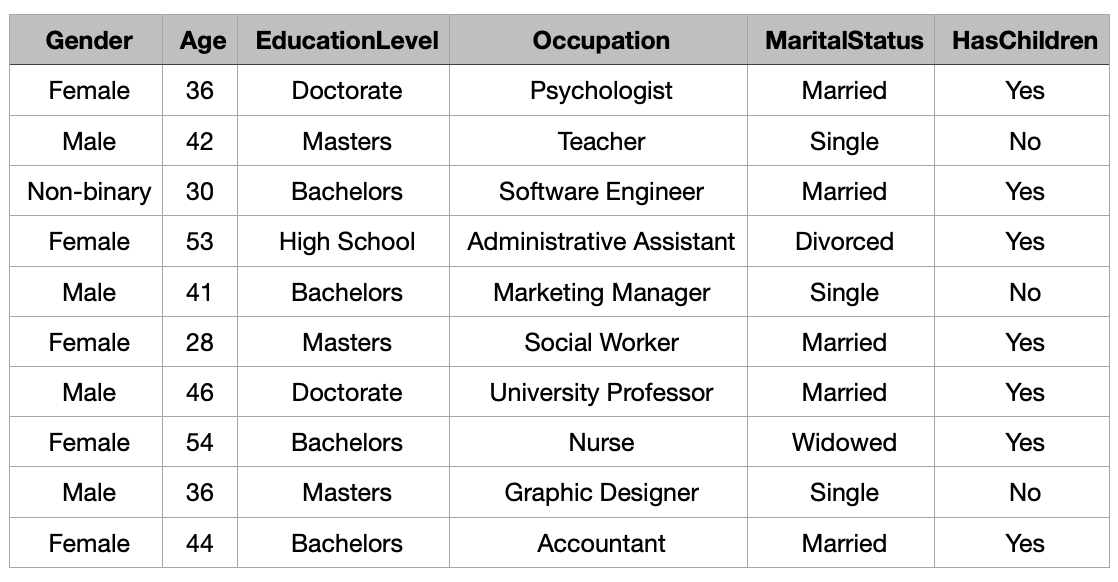}
    
    \vspace{0.2cm}

\begin{lstlisting}[linewidth=\textwidth]
 prefix dsv:<https://w3id.org/dsv-ontology#>.
 prefix dcterms: <http://purl.org/dc/terms/>.
 prefix rdfs: <http://www.w3.org/2000/01/rdf-schema#> .
 prefix dbpedia: <http://dbpedia.org/ontology/> .

<http://metaUnionSearch/datasets/PsychologyUEA3GE8N> a dsv:Dataset ;
    dcterms:subject "psychology" ;
    dcterms:title "Psychology_UEA3GE8N.csv" ;
    dsv:datasetSchema [ dsv:column 
    <http://metaUnionSearch/datasets/PsychologyUEA3GE8N/column/Gender>,
    <http://metaUnionSearch/datasets/PsychologyUEA3GE8N/column/Age>,
    <http://metaUnionSearch/datasets/PsychologyUEA3GE8N/column/EducationLevel>,
    <http://metaUnionSearch/datasets/PsychologyUEA3GE8N/column/Occupation>,
    <http://metaUnionSearch/datasets/PsychologyUEA3GE8N/column/MaritalStatus>,
    <http://metaUnionSearch/datasets/PsychologyUEA3GE8N/column/HasChildren>. ] .

<http://metaUnionSearch/datasets/PsychologyUEA3GE8N/column/Gender> a dsv:Column ;
    rdfs:label " Gender" ;
    dcterms:type "gender" ;
    dsv:columnProperty dbpedia:gender . 
    
<http://metaUnionSearch/datasets/PsychologyUEA3GE8N/column/Age> a dsv:Column ;
    rdfs:label "Age" ;
    dcterms:type "age" ;
    dsv:columnProperty dbpedia:age .

<http://metaUnionSearch/datasets/PsychologyUEA3GE8N/column/EducationLevel> a dsv:Column ;
    rdfs:label "EducationLevel" ;
    dcterms:type "education" ;
    dsv:columnProperty dbpedia:education .

<http://metaUnionSearch/datasets/PsychologyUEA3GE8N/column/Occupation> a dsv:Column ;
    rdfs:label "Occupation" ;
    dcterms:type "position" ;
    dsv:columnProperty dbpedia:occupation .
    
<http://metaUnionSearch/datasets/PsychologyUEA3GE8N/column/MaritalStatus> a dsv:Column ;
    rdfs:label "MaritalStatus" ;
    dcterms:type "status" ;
    dsv:columnProperty dbpedia:spouse .
    
<http://metaUnionSearch/datasets/PsychologyUEA3GE8N/column/HasChildren> a dsv:Column ;
    rdfs:label "HasChildren" ;
    dcterms:type "status" ;
    dsv:columnProperty dbpedia:child .
\end{lstlisting}
\vspace{0.1cm}
    \caption{Excerpt of a CSV dataset from the ALT-gen benchmark, with its corresponding metadata. In the metadata we can see the enrichments, added with the dcterms:type property for the semantic data types, and with the dsv:columnProperty property for the DBpedia properties.}
    \label{fig:metadata-example}
\end{figure}

In the test set, the total number of triples before enrichment was 21,626, and after enrichment with semantic data types and DBpedia properties, it increased to 31,775. In the evaluation set, the number of triples before enrichment was 32,126, and after enrichment, it grew to 47,401. The \textit{base} triples (i.e. without enrichments) refer to those extracted directly from the CSV files, while the \textit{enriched} triples include the additional column-level metadata properties, such as \verb|dcterms:type| for semantic data types and \verb|dsv:columnProperty| for DBpedia properties. Figure \ref{fig:metadata-example},below, shows an excerpt of a dataset from the ALT-gen benchmark, along with its metadata.

\subsection{Metadata Union Search}
Here we present the results of the Metadata Union Search (MUS), evaluating the unionability of tables solely using metadata information. To compute the unionability for the evaluation set, we first needed to calculate a threshold (MUS\textsubscript{t}) for the cosine similarity between the query and candidate tables of the test set. (MUS\textsubscript{t}) is essentially a cosine similarity score, which we set as the threshold for determining unionability. The closer it is to 1, the more strict the requirement for the datasets to be considered unionable, as a cosine similarity of 1 represents exact similarity between the datasets. (MUS\textsubscript{t}) was determined by comparing the cosine similarity scores of the test set against the ground truth from the ALT-gen benchmark, and then applying the Youden's statistic. Once the threshold was calculated, we applied it to the evaluation set, considering query and candidate pairs to be unionable if their cosine similarity exceeded the threshold. 

The MUS\textsubscript{t}) observed in the ``Topic Dependent'' (TD) scenario were generally lower, ranging from 0.63 to 0.69. For the ``Topic Guided'' (TG) scenario, instead, the threshold values were higher, ranging from 0.79 to 0.84. A possible explanation for the lower MUS\textsubscript{t}) in the TD scenario is that unionability is computed only between candidates and queries from the same topic. On the other hand, the TG scenario involves the comparison of cross-topics query and candidate pairs, which increases the likelihood of mistakes. As a result, Youden's statistic returned a higher threshold to ensure that the predictions for unionability are accurate and to prevent false positives.

Additionally, we calculated the overall accuracy and precision for each scenario and each enrichment setting. In the TD scenario, we found that metadata with DBpedia enrichment had the highest accuracy and precision. In the TG scenario, the highest accuracy and precision were achieved without semantic enrichment (referred to as \textit{base}). We also observed that the thresholds were highest in these respective configurations. However, we do not have enough data to conclude any statistical significance between these findings. As above, a possible reason for this difference can be related to the fact that in the TG scenario could require stricter MUS\textsubscript{t}) to avoid errors, while int the TD scenario might be easier to find unionable tables. 

Furthermore, we computed accuracy and precision for both unionable (1) and non-unionable (0) predictions. This distinction is particularly important because our main objective is to achieve strong performance for unionable predictions. Separating these categories allows for a better understanding of the ability of our method to make accurate predictions, especially about unionable tables. All the results discussed are shown in Table \ref{tab:metrics-unionable-ununionable}.

\begin{table}[]
\centering
\begin{tabular}{p{1cm}p{2cm}>{\centering\arraybackslash}m{2cm}*{6}{>{\centering\arraybackslash}p{1cm}}}
\toprule
 &  & \multirow{2}{*}{\textbf{MUS\textsubscript{t}}} & \multicolumn{3}{c}{Accuracy} & \multicolumn{3}{c}{Precision} 
 \\\arrayrulecolor{black}\cmidrule(lr){4-6}\cmidrule(lr){7-9}
 &  &  & $\bullet$ & 1 & 0 & $\bullet$ & 1 & 0 \\\midrule
\multirow{4}{*}{TD}
 & base     & 0.67  & 0.71  & \textbf{0.82} & 0.60  & 0.67  & 0.67  & 0.77 \\
 & dtypes   & 0.63  & 0.65  & 0.77  & 0.53  & 0.62  & 0.62  & 0.70 \\
 & dbpedia  & 0.69 & \textbf{0.72} & 0.57  & \textbf{0.87}  & \textbf{0.82}  & \textbf{0.82}  & 0.67 \\
 & dtypes+dbpedia & 0.66 & 0.69 & 0.66 & 0.72 & 0.70 & 0.70 & 0.68 \\
 \arrayrulecolor[rgb]{0.502,0.502,0.502}\hline
\multirow{4}{*}{TG}
 & base & 0.84 & \textbf{0.70} & 0.81 & \textbf{0.59} & \textbf{0.66} & \textbf{0.66} & \textbf{0.76} \\
 & dtypes & 0.82 & 0.63 & 0.76 & 0.50 & 0.60 & 0.60 & 0.68 \\
 & dbpedia & 0.81 & 0.69 & 0.81 & 0.56 & 0.65 & 0.65 & 0.75 \\
 & dtypes+dbpedia & 0.79 & 0.61 & 0.81 & 0.42 & 0.58 & 0.58 & 0.68 \\
 \arrayrulecolor{black}\bottomrule
\end{tabular}
\vspace{0.1cm}
\caption{Evaluation of unionability for Metadata Union Search (MUS) under different enrichment settings. The table presents accuracy and precision metrics for ``unionable'' (1), ``non-unionable'' (0) and overall ($\bullet$) predictions across the ``Topic Dependent'' (TD) and ``Topic Guided'' (TG) scenarios. Results are shown for various enrichment configurations, including base (without enrichment), semantic data types enrichment (dtypes), DBpedia properties enrichment, and a combination of data types and DBpedia. The highest accuracy and precision values for each metric are highlighted in bold.}
\label{tab:metrics-unionable-ununionable}
\end{table}

\subsection{Comparison of Unionability Methods}
In this section, we present a comparison of our Metadata Union Search (MUS) method against other available methods, using the ALT-gen benchmark. The results, displayed in Table \ref{tab:comparison-benchmarks}, include performance data reported in the ALT-gen benchmark paper for the other methods. Following the approach outlined in their work, the comparison is based on three evaluation metrics: Mean Average Precision at k (MAP@k), Precision at k (P@k), and Recall at k (R@k), all evaluated at \verb|k = 10|. These metrics assess the effectiveness of unionability predictions, on the top 10 results for each method. The compared methods include D\textsuperscript{3}L\cite{bogatu2020dataset}, SANTOS\cite{khatiwada2023santos} and Starmie\cite{fan2022semantics}.

Our results show that MUS consistently outperforms both D\textsuperscript{3}L and SANTOS across all metrics and for all variants. For the Starmie methods, the MAP@k metric is somewhat comparable, with both achieving a MAP@k of 0.71 in the \textit{base} setting. However, MUS consistently outperforms Starmie in terms of P@k and R@k across all variants. We can also see that the different settings (i.e. base, dtypes, dbpedia) leads to some variation in performance, with the best results consistently observed in the base settings for both TD (topic-dependent) and TG (topic-guided) scenarios. The performance in the TG scenario also shows that MUS with the \textit{base} setting reaches strong P@k and R@k scores. The difference in performance between TD and TG scenarios highlights that our method can be suitable for different contexts, allowing for flexible table unionability prediction with or without the availability of topic filtering.

Overall, these results shows that while MUS does not always achieve the highest MAP@k values when compared to Starmie, it consistently outperforms other methods, especially in terms of P@k and R@k. Based on this outcome, we highlight that MUS performs competitively while being based solely on metadata information. Furthermore, the flexibility of MUS, with various enrichment settings, provides additional opportunities for improving performance depending on the specific scenario, making it a flexible and efficient method for unionability prediction, even when only the metadata is available.

\begin{table}
\centering
\arrayrulecolor[rgb]{0.502,0.502,0.502}
\begin{tabular}{p{6.5cm} >{\centering\arraybackslash}p{1.5cm} >{\centering\arraybackslash}p{1.5cm} >{\centering\arraybackslash}p{1.5cm}}
\arrayrulecolor{black}\cmidrule[\heavyrulewidth]{1-1}\cmidrule[\heavyrulewidth]{2-2}\cmidrule[\heavyrulewidth]{3-3}\cmidrule[\heavyrulewidth]{4-4}
\multicolumn{1}{p{3cm}}{\textbf{Method}} & \textit{MAP@k} & \textit{P@k} & \textit{R@k} \\ 
\midrule
D\textsuperscript{3}L & 0.15 & 0.13 & 0.13 \\
SANTOS & 0.43 & 0.27 & 0.27 \\ 
Starmie & 0.71 & 0.56 & 0.56 \\ 
Starmie-Vicuna\textsubscript{Zero} & 0.65 & 0.48 & 0.48 \\ 
Starmie-Vicuna\textsubscript{Optim} & 0.71 & 0.56 & 0.56 \\ 
\arrayrulecolor[rgb]{0.502,0.502,0.502}\hline\hline
MUS - base\textsubscript{(TD)} & 0.69 & 0.72 & 0.72 \\
MUS - dtypes\textsubscript{(TD)} & 0.63 & 0.67 & 0.67 \\
MUS - dbpedia\textsubscript{(TD)} & 0.55 & 0.56 & 0.56 \\
MUS - dtypes+dbpedia\textsubscript{(TD)} & 0.56 & 0.60 & 0.60 \\
\arrayrulecolor[rgb]{0.502,0.502,0.502}\hline
MUS - base\textsubscript{(TG)} & 0.69 & 0.72 & 0.72 \\
MUS - dtypes\textsubscript{(TG)} & 0.60 & 0.63 & 0.63 \\
MUS - dbpedia\textsubscript{(TG)} & 0.65 & 0.69 & 0.68 \\
MUS - dtypes+dbpedia\textsubscript{(TG)} & 0.62 & 0.67 & 0.67 \\
\arrayrulecolor{black}\bottomrule
\end{tabular}
\vspace{0.1cm}
\caption{Comparison of the Metadata Union Search (MUS) method with other existing methods (D\textsuperscript{3}L, SANTOS and Starmie) on the basis of MAP@k, P@k, and R@k metrics. The results are shown for both the ``Topic Dependent'' (TD) and ``Topic Guided'' (TG) scenarios, with different enrichment settings.}
\label{tab:comparison-benchmarks}
\end{table}

\section{Discussion}
In this study, we introduced Metadata Union Search (MUS), a novel method for computing table unionability using solely metadata, and evaluated it against state-of-the-art methods using the ALT-gen benchmark. Our results show that MUS achieves competitive performance across both topic-dependent (TD) and topic-guided (TG) scenarios. By leveraging semantic technologies, MUS demonstrates its capability to predict unionability with high accuracy scores, even in the absence of the underlying data. These results highlight the potential of metadata-driven approaches to bridge the gap between open data and restricted access data.

Our analysis also reveals that MUS can be adapted to different contexts. While semantic enrichment improves performance in certain scenarios, the base configuration without enrichments consistently shows reliable and robust predictions. Comparisons with existing methods such as D\textsuperscript{3}L, SANTOS, and Starmie, particularly in P@k and R@k metrics, shows that MUS is a suitable solution for metadata-driven table unionability tasks.

\subsubsection{Future Work}
While our method provides a foundation for metadata-driven unionability, there are several possibilities for future research. First, the integration of domain-specific ontologies and knowledge repositories could further enhance the accuracy of unionability predictions. Another option could be the incorporation of hierarchical metadata annotations, which may allow for more complex retrieval tasks. Finally, validating the approach on additional benchmarks and real-world datasets beyond ALT-gen would help generalize the findings and showcase the broader potential of metadata-driven methods for table unionability.

\begin{credits}
\subsubsection{Declarations} 
We acknowledge that ChatGPT was utilized to generate and debug part of the python and latex code utilised in this work. 
Conflict of Interest: this work is funded by the Netherlands Organization of Scientific Research (NWO), ODISSEI Roadmap project: 184.035.014. 
\end{credits}
%
%
%
\bibliographystyle{apalike}
\bibliography{main.bib}

\end{document}